\documentclass[a4paper]{jpconf}

\usepackage{graphicx}
\usepackage{geometry}
\usepackage{color}
\usepackage[varg]{txfonts}  

\begin{document}

\title{Transport Theory from the Nambu-Jona-Lasinio Lagrangian}

\author{ R. Marty$^{a,b,c}$, J. M. Torres-Rincon$^a$, E. Bratkovskaya$^{b,c}$  and J. Aichelin$^a$\footnote{invited lecturer}}

\address{$^a$ SUBATECH, Laboratoire de Physique Subatomique et des Technologies Associ\'ees,\\
Universit\'e de Nantes - IN2P3/CNRS - Ecole des Mines de Nantes\\
4 rue Alfred Kastler, F-44072 Nantes, Cedex 03, France\\
$^b$ Institute for Theoretical Physics, JWG Universit\"{a}t, Frankfurt/M, Germany\\
$^c$ Frankfurt Institute for Advanced Studies, JWG Universit\"{a}t, Frankfurt/M, Germany}
\ead{aichelin@in2p3.fr}
\date{}  


\begin{abstract}
Starting from the (Polyakov-) Nambu-Jona-Lasinio Lagrangian, (P)NJL, we formulate a transport theory which allows for describing the expansion of a quark-antiquark plasma and the subsequent transition to the hadronic world ---without adding any new parameter to the standard (P)NJL approach, whose parameters are fixed to vacuum physics. This transport theory can be used to describe ultrarelativistic heavy-ion reaction data as well as to study the (first-order) phase transition during the expansion of the plasma. (P)NJL predicts such a phase transition for finite chemical potentials. In this contribution we give an outline of the necessary steps to obtain such a transport theory and present first results. 
\end{abstract}

\section{Introduction}

The Nambu-Jona-Lasinio (NJL) Lagrangian shares the symmetries of the Lagrangian of Quantum Chromo Dynamics (QCD). Since symmetries are nowadays considered as the most important properties of a Lagrangian, it has been used for many studies of ground-state properties of strongly interaction matter at zero temperature, as well as at finite temperatures and finite chemical potentials. This includes the studies of the hadron-parton transition and of the color-flavor locking. It has also been used very successfully in the past to study meson and baryon properties as functions of the temperature and of the baryon density like hadron masses, widths and decay constants. The NJL Lagrangian contains quark fields but no gluon fields. The quarks interact by a four-point interaction.  Recently the standard NJL Lagrangian has been extended to the Polyakov NJL (PNJL) which includes gluons on a mean-field level~\cite{Ratti:2005jh}. 

Having a four-fermion interaction the (P)NJL Lagrangian is not renormalizable. It does also not confine the quarks into colorless objects. Therefore it is a challenge to describe the transition of a plasma of quarks and antiquarks to hadrons with such an approach. Usually the calculations using the (P)NJL Lagrangian have been performed at the mean-field level. Going calculationally beyond the mean field by including systematically the fluctuations, the meson contribution to the equation of state could be identified~\cite{Zhuang:1994dw,Hufner:1994ma}. 

In these short proceedings we cannot cover all these developments and therefore, we refer to the recent reviews for details~\cite{Klevansky:1992qe,Vogl:1991qt,Ebert:1994mf,Buballa:2003qv}. Here we concentrate on the steps which are necessary to develop a transport theory from the (P)NJL Lagrangian. After having introduced the Lagrangian in section 2, discuss in section 3 how this approach is extended to finite temperature and density and present the calculation of the masses of quarks, mesons and baryons at finite temperature. Section 4 contains the outline of the covariant transport theory which can be built with the help of the NJL Lagrangian. In section 5 we present first results of this transport theory in comparison with experimental data of ultrarelativistic heavy-ion collisions. Section 6, finally, concludes this lecture with a summary.
   
\section{The starting point: The (P)NJL Lagrangian}

The thermal properties of QCD, the theory of strong interactions, can presently only be calculated computationally (``lattice QCD''), and this can be done exclusively for the case of grand canonical  systems with zero chemical potential. This allows for the determination of the equation of state and other thermal properties, but already the thermal properties of the constituents of the theory are not precisely known. In addition, only few of the transport coefficients which can shed light on the evolution of a system out of (but close to) equilibrium are yet calculated on lattice-QCD with still considerable error bars.

If one wants to study the dynamics of strongly interacting matter, one has therefore to rely on effective theories. One of these effective theories is the NJL model based on the Lagrangian (in the SU(3) version)
\begin{eqnarray}
\label{eq:Lex}
\mathbf{ \mathcal{L}_{NJL}}&=& \sum_i \bar \psi_i (i\gamma_\mu\partial^\mu-\hat m_i)\psi_i   - G^2_c \sum_{ijkl} \ [ \bar{\psi}\textcolor{blue}{_i} \textcolor{green}{\gamma^\mu} \textcolor{red}{T^{a'}} \textcolor{blue}{\delta_{ij}}\psi \textcolor{blue}{_j} ] \  \ [ \bar{\psi}\textcolor{blue}{_k} \textcolor{green}{\gamma_\mu} \textcolor{red}{T^{a'}} \textcolor{blue}{\delta_{kl}} \psi \textcolor{blue}{_l}] \nonumber \\
&+&  H \det_{ij} \ [\bar{\psi}_i (1-\gamma_5) \psi_j] - H \det_{ij} \ [\bar{\psi}_i (1+ \gamma_5) \psi_j ]
\end{eqnarray}
where $\psi_i $ are spinors of quarks of flavor $i$. The second term is a four-point interaction in Dirac (green), color (red) and flavor (blue) spaces. This term can be related to perturbative QCD by realizing that it describes in  Born approximation the interaction of two quark currents mediated by a gluon whose mass is large as compared to the momentum transfer \cite{Ebert:1994mf}. This term is followed by the 't Hooft determinant which gives rise to the $U_A(1)$ symmetry breaking.  $T^{a'}$ are the color SU(3) Gell-Mann matrices with $ \textrm{tr}_c \  (T^{a'} T^{b'}) = 2\delta^{a'b'}$. The mass of the strange quarks is different from that of up and down quarks which are considered as degenerated.  Being an effective model of QCD the NJL Lagrangian respects the symmetries of the underlying theory, in particular the $U_V(1) \otimes SU_V(N_f) \otimes SU_A(N_f)$ global symmetries of the massless QCD Lagrangian. The $U_V(1)$ symmetry leads to the baryon number conservation, while the chiral symmetry $SU_V(N_f) \otimes SU_A(N_f)$ is spontaneously broken down to $SU_V(N_f)$ at low temperatures. The $U_A(1)$ symmetry is broken by the axial anomaly. 


After performing a Fierz transformation~\cite{Klevansky:1992qe,Buballa:2003qv,Torres-Rincon:2015rma} this Lagrangian can be reexpressed in a convenient way to describe the $qq$, $\bar{q}\bar{q}$ and $\bar{q}q$ interactions. The $\bar{q} q$ sector of the effective theory will allow us to describe mesons. For instance, the pseudoscalar sector of the interacting Lagrangian (\ref{eq:Lex}) reads (all repeated indices are to be summed)
\begin{equation}
\label{eq:fierz1}
{\cal L}_{\bar{q}q} = G \ (\bar{\psi}_i \ i\gamma_5 \ \tau^a_{ij} \ \psi_j) 
(\bar{\psi}_k \ i\gamma_5 \ \tau^a_{kl} \ \psi_l) \ ,
\end{equation}
where $a=1,...,N_f^2-1$ and $G$ is a coupling constant. We will take $G$ as a free parameter to be fixed by comparing our results with the experimental hadron spectrum. The flavor Gell-Mann matrices $\tau^a$ follow the normalization
\begin{equation}
\textrm{ tr } (\tau^a \tau^b) = 2 \delta^{ab} \ .
\end{equation}

The Fierz transformation  produces diquarks as well. We detail here the Lagrangian describing the scalar diquark sector
\begin{equation}
\label{eq:lagdiq}
{\cal L}_{qq} = G_{DIQ} \ (\bar{\psi} i\gamma_5 C \tau^A T^{A'} \bar{\psi}^T) 
(\psi^T C i \gamma_5 \tau^{A} T^{A'} \psi) \ ,
\end{equation}
and the one for the axial diquark sector
\begin{equation}
\label{eq:lagdiq1}
{\cal L}_{qq} = G_{DIQ,V} \ (\bar{\psi} \gamma^\mu C \tau^S T^{A'} \bar{\psi}^T) 
(\psi^T C  \gamma_\mu \tau^{S} T^{A'} \psi) \ , 
\end{equation}
where $G_{DIQ}$ and $G_{DIQ,V}$ are coupling constants (related to the original $G_c^2$ but taken here as free parameters) and $C=i\gamma_0 \gamma_2$ represents the charge-conjugation operator. Finally, we have denoted by $\tau^A$ and $\tau^S$ the antisymmetric and symmetric SU(3) flavor matrices, respectively; and by $T^{A'}$ the antisymmetric color matrices. In particular, the presence of the latter reflects that the diquarks cannot be color singlets. 
 
The (P)NJL Lagrangian has several parameters which have to be fixed. In our calculation we take those of Table~\ref{tab:param}.
\begin{table*}
\begin{center}
\begin{tabular}{|c|c|c|c|c|c|c|c|}
\hline
Parameter & $m_{q0}$ & $m_{s0}$ & $\Lambda$  &$G$ & $H$ & $G_{DIQ}$ & $G_{DIQ,V}$  \\
\hline
Value & 5.5 MeV & 134 MeV & 569 MeV& $2.3/\Lambda^2$ & $11/\Lambda^5$& $1.56 \ G$& $-0.639 \ G_{DIQ}$ \\
\hline
\hline
Parameter & $a_0$ & $a_1$ & $a_2$ & $a_3$ & $b_3$ & $b_4$ & $T_0$ \\ 
\hline
Value  & 6.75 & -1.95 & 2.625 & -7.44 & 0.75 & 7.5 & 190 MeV \\ 
\hline
\end{tabular}
\caption{\label{tab:param} Parameters of the NJL and PNJL model used in this study. In the isospin limit
we have $m_{q0}=m_{u0}=m_{d0}$.}
\end{center}
\end{table*}

$\Lambda$ is the momentum cut-off which has to be applied in the loop integrations. Using the parameter set in Table~\ref{tab:param} we obtain at $T=0$: the light-quark condensate $\langle \bar{\psi}_u \psi_u\rangle=-(241.3$ MeV$)^3$, the pion decay constant $f_\pi=92.2$ MeV, the pion mass $m_\pi=134.8$ MeV, the kaon mass $m_K=492.1$ MeV, the $\eta-\eta'$ mass splitting of $475.5$ MeV, the proton mass $932.0$ MeV and the $\Delta$ baryon mass $1221.4$ MeV. 

This means that all our parameters are fixed by vacuum hadron masses and decay constants. We would like to stress, and this is the great advantage of the (P)NJL model, that no other parameters are needed to describe hadron masses at finite temperature and density, the pressure, trace anomaly and entropy density as a function of $T$ and $\mu$, the cross sections for scattering among quarks and antiquarks and the hadronization cross sections. Finally the whole transport approach and its prediction for the observables do not need more than these parameters.  

\section{Finite-temperature physics}

To calculate the hadron properties at  finite temperature, we use the imaginary time formalism with the prescription
\begin{equation}
  \int \frac{d^4 k}{(2\pi)^4} \to i T \sum_{n \in \mathbb{Z}} \int \frac{d^3 k}{(2\pi)^3} \ , 
\end{equation}
with $T$ the temperature and $k^0 \rightarrow i \omega_n$ the fermionic Matsubara frequencies $i\omega_n =i \pi T (2n+1)$.

To account for the finite baryonic density we can introduce a quark chemical potential by adding to the Lagrangian the term
\begin{equation}
  {\cal L}_\mu = \sum_{ij} \bar{\psi}_i \ \mu_{ij} \gamma_0 \ \psi_j \ ,
\end{equation}
where $\mu_{ij} = \textrm{ diag } (\mu_u,\mu_d,\mu_s)$ contains the quark chemical potentials (which can be alternatively expressed in terms of the baryon, charge, and strangeness chemical potentials, $\mu_B,\mu_Q,\mu_S$). In this work we will restrict ourselves to a vanishing chemical potential $\mu_u=\mu_d=\mu_s=0$.

In the PNJL model an effective potential ${\cal U} (\Phi,\bar{\Phi},T)$, is added to the effective NJL Lagrangian ${\cal L} \rightarrow {\cal L} - {\cal U}$. ${\cal U}$ is a function of the Polyakov loop $\Phi$
\begin{equation}
  \Phi = \frac{1}{N_c} \textrm{ tr}_c \langle L \rangle \ \quad ; \quad
  L ({\bf x})= {\cal P} \exp \left( i \int_0^\beta d\tau A_4(\tau,{\bf x}) \right) \ , 
\end{equation}
and its complex conjugate, which are taken to be homogeneous fields. The form of the effective potential is inspired by the $\mathbb{Z}_3$ center symmetry~\cite{Ratti:2005jh}
\begin{equation}
  \frac{{\cal U} (T,\Phi,\bar{\Phi})}{T^4} = - \frac{b_2(T)}{2} \bar{\Phi} \Phi - \frac{b_3}{6}\left( \Phi^3 + \bar{\Phi}^3 \right) + \frac{b_4}{4} \left(  \bar{\Phi} \Phi \right)^2 \ , 
\end{equation}
with
\begin{equation}
  b_2(T) = a_0 + a_1 \frac{T_0}{T} + a_2  \left( \frac{T_0}{T} \right)^2 + a_3  \left( \frac{T_0}{T} \right)^3 \ .
\end{equation}
The parameters $a_0,a_1,a_2,a_3,b_3,b_4$ and $T_0$ are fitted from the pure-gauge lattice-QCD equation of state (with a reduction of $T_0$ following the arguments in~\cite{Schaefer:2007pw}) and can be found in Table~\ref{tab:param}.

A detailed calculation shows \cite{Hansen:2006ee,Costa:2008dp} that including the Polyakov loop leads to the replacement of the Fermi distribution function, which appears in the finite temperature calculation using the NJL Lagrangian,  by
\begin{eqnarray}
f_\Phi^+ (E_i) &=& \frac{ ( \Phi + 2 \bar{\Phi} e^{- E_i/T} ) e^{- E_i/T} + e^{-3 E_i/T}}%
{1+3( \Phi + \bar{\Phi} e^{- E_i/T}) e^{- E_i/T } + e^{-3 E_i/T}} \ , \\
f_\Phi^- (E_i) &=& \frac{ ( \bar{\Phi} + 2 \Phi e^{- E_i/T} ) e^{- E_i/T} + e^{-3 E_i/T}}%
{1+3( \bar{\Phi} + \Phi e^{- E_i/T}) e^{- E_i/T } + e^{-3 E_i/T}} \ .
\end{eqnarray}
where + and - refer to a negative (positive) chemical potential $(E+\mu)((E-\mu))$ and $\Phi (\bar \Phi)$ is obtained by minimizing the PNJL grand-canonical potential,
\begin{eqnarray}
\Omega_{PNJL} (\Phi,\bar{\Phi},m_i,T) &=& {\cal U}(T,\Phi,\bar{\Phi}) + 2G \sum_i \langle \bar{\psi}_i \psi_i \rangle^2 
 - 4H \prod_i  \langle \bar{\psi}_i \psi_i \rangle  - 2N_c \sum_i \int \frac{d^3 k}{(2\pi)^3} E_i \nonumber \\
&-& 2 T \sum_i \int \frac{d^3 k}{(2\pi)^3} \left[ \textrm{ tr}_c \log \left(1+L e^{-E_i/T} \right) 
+  \textrm{ tr}_c \log \left(1+L^\dag e^{-E_i/T} \right)  \right] \ , 
\end{eqnarray}
(with $E_i=\sqrt{k^2 + m_i^2}$) with respect to  $\Phi ({\bar \Phi})$,
\begin{equation}
\label{eq:min}
  \frac{\partial \Omega_{PNJL}}{\partial \Phi} =0 \ , \quad \frac{\partial \Omega_{PNJL}}{\partial \bar{\Phi}} =0.
\end{equation}
The minimization of the PNJL grand-canonical potential with respect to $\langle \bar{\psi}_i \psi_i \rangle$ gives the PNJL quark condensate
\begin{equation}
\label{eq:condenpnjl}
  \langle \bar{\psi}_i \psi_i \rangle = - 2N_c \int  \frac{d^3 k}{(2\pi)^3} \frac{m_i}{E_i} \left[1 - f_\Phi^+(E_i) - f_\Phi^- (E_i)\right]  \ .
\end{equation}

\subsection{Quark and meson masses}

\begin{figure*}[htp]
  \begin{center}
    \includegraphics[width=0.9\textwidth]{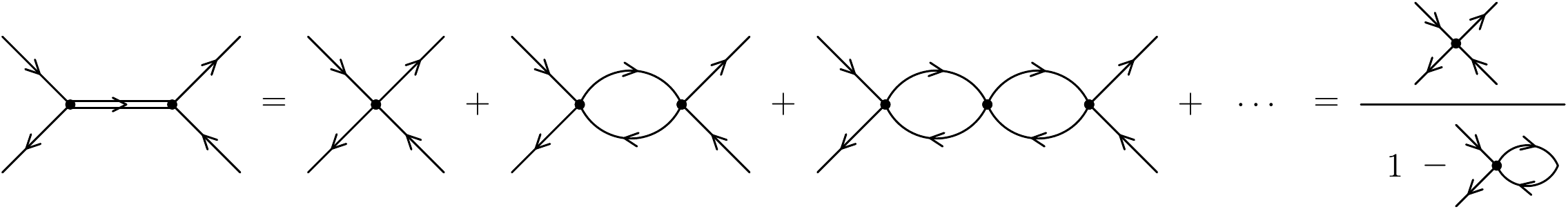}
    \caption{Effective interaction between a quark and an antiquark in the Random Phase Approximation.} 
    \label{bethes}
  \end{center} 
\end{figure*}

In this model~\cite{Torres-Rincon:2015rma} chiral symmetry restoration --expressed by the decreasing quark condensate (Eq.~\ref{eq:condenpnjl}) with temperature-- lowers the quark masses with increasing temperature
\begin{equation}
  M_i = m_{i0}- 4 G \langle \bar{\psi}_i \psi_i \rangle - 2H \langle \bar{\psi}_j \psi_j \rangle \langle \bar{\psi}_k \psi_k \rangle
\label{mesonm}
\end{equation}
with $i,j,k = u,d,s$ (to be fixed in cyclic order). Mesons are obtained as poles of the quark-antiquark scattering amplitude in the complex-energy plane. The scattering amplitude is the solution of the Bethe-Salpeter equation, formally displayed in Fig.~\ref{bethes}
$q_i+\bar q_j \to q_k+\bar q_l $ (where the indices refer to the quark flavor) and which is solved in the random-phase or Brueckner approximation (which are the same for local four-point interactions)
\begin{equation}
T^{ab}_{i \bar{j},m \bar{n}} (p^2) = {\cal K}^{ab}_{i \bar{j},m \bar{n}} + i \int \frac{d^4 k}{(2 \pi)^4}
{\cal K}^{ac}_{i \bar{j}, p \bar{q}} \  S_{p} \left( k+ \frac{p}{2} \right) \ S_{ \bar{q}}  \left( k-\frac{p}{2} \right)
\ T^{cb}_{p \bar{q},m \bar{n}} (p^2) \ ,
\end{equation}
where $a,b$ denotes the meson flavor channel. The kernel ${\cal K}$ reads
\begin{equation}
{\cal K}^{ab}_{i\bar{j},m\bar{n}} = \Omega^a_{i \bar{j}} \ 2 K^{ab} \ \bar{\Omega}^b_{\bar{n}m} \ ,
\end{equation}
with the vertex factors containing color, flavor and spin matrices
\begin{equation}
\Omega^a_{i \bar{j}}= \left( \mathbb{I}_{\textrm{color}} \otimes \tau_{i \bar{j}}^a \otimes \Gamma \right)\ ,
\end{equation}
as well as a combinatorial factor of 2. The Dirac structure --whose indices we have omitted in the BS equation-- can be chosen to be $\Gamma=\{1,i\gamma_5, \gamma^\mu, \gamma_5 \gamma^\mu \}$ for scalar, pseudoscalar, vector, and axial-vector mesons, respectively.
\begin{figure*}[htp]
  \begin{center}
    \includegraphics[width=0.35\textwidth]{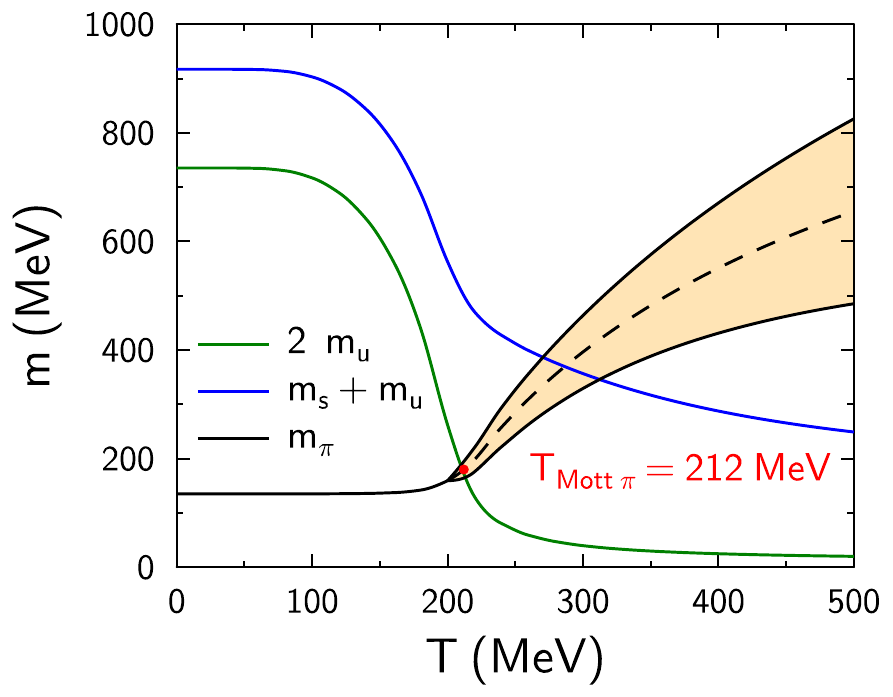}
    \includegraphics[width=0.35\textwidth]{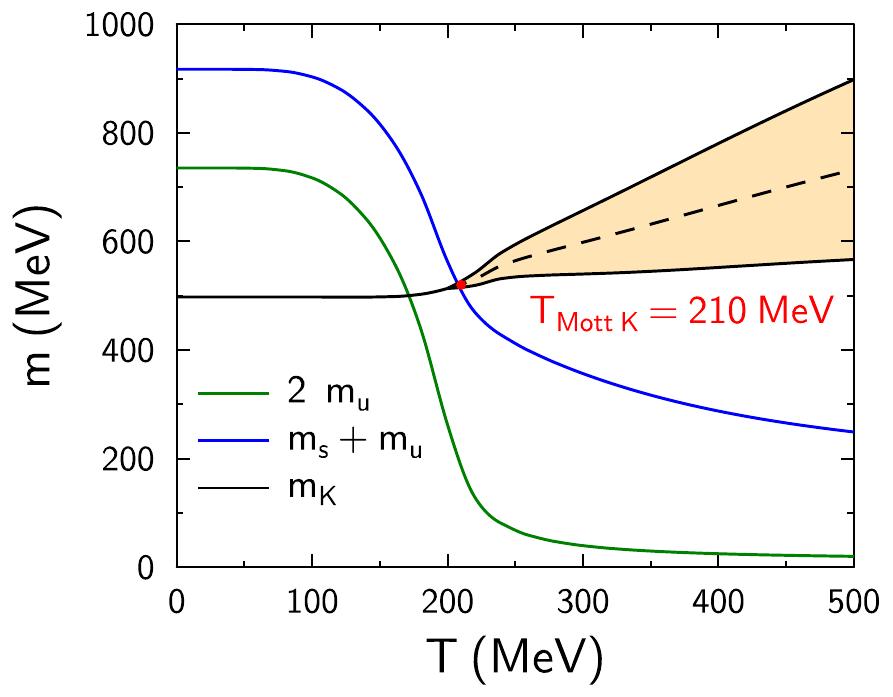}
    \caption{\label{fig:Pmesons} Left panel: Mass of $\pi$ meson and of light quarks, right panel: Mass of K meson and light and heavy quark as a function of the temperature.} 
  \end{center}
\end{figure*}
Figure~\ref{fig:Pmesons} displays the masses of mesons and their constituent quarks as a function of the temperature for $\mu_i=0$. We see that the quark masses (Eq.~\ref{mesonm}) are about the constituent quark masses ($\approx M_{proton}/3.$) at low temperature and decrease as a function of the temperature due to the melting of the quark condensate. At very large temperature they are very close to the bare mass. The pseudoscalar mesons are the Goldstone bosons that result from the spontaneous breakdown of the chiral-flavor symmetries of QCD, and are therefore little influenced by the temperature. At low temperature their masses are lower than that of the two constituent quarks, and therefore they are the thermodynamically relevant degrees of freedom. Beyond the Mott temperature the sum of the constituent quark masses is lower than the mass of the meson which develops a width because it may decay into its constituents. The Mott temperatures of $K$'s and $\pi$'s are not identical but close by. Beyond the Mott temperature the quarks are the thermodynamically relevant quantities and a plasma of quarks and antiquarks is formed.

\subsection{Baryons}

Diquarks can form --together with a third quark-- color neutral baryons~\cite{Torres-Rincon:2015rma}.  For $N_f=3$ we will consider both the octet and decuplet flavor representations of baryons. Scalar diquarks (those belonging to flavor $\bar{{\bf 3}}$ representation) and axial diquarks (${\bf 6}$) will be used to build up the baryon octet and decuplet states, respectively, according to the decomposition,
\begin{equation}
  {\bf 3} \otimes ( \bar{ { \bf 3}} \oplus {\bf 6}) = ({\bf 1} \oplus {\bf 8}) \oplus ({\bf 8} \oplus {\bf 10}) \ .
\end{equation}

The starting point to describe baryons is the Fadeev equation~\cite{Buck:1992wz,Ishii:1995bu}:
\begin{equation}
\label{eq:fadeev}
  \left.  X_j^{\bar{j} ,\alpha} (P^2,q) - \int \frac{d^4 k}{(2\pi)^4} 
  L_{j k}^{\bar{j} \bar{k}, \alpha \beta}(P^2,q,k) X_{k}^{\bar{k} ,\beta} (P^2,k)\right|_{P^2=M^2_B}= 0 \ ,
\end{equation}
where the baryon wave function is denoted by $X_j^{\bar{j}, \alpha}$ and it carries a quark index ($j$), diquark index ($\bar{j}$), and a possible spin index $\alpha$.

\begin{figure*}[htp]
  \begin{center}
    \includegraphics[width=0.4\textwidth]{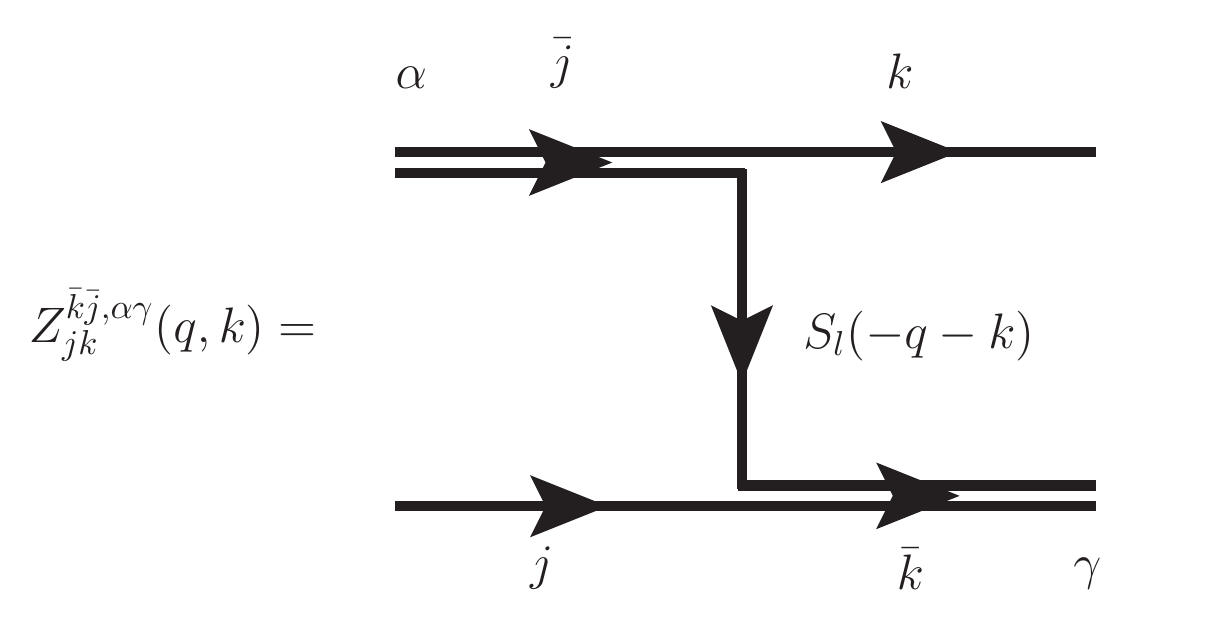}
    \includegraphics[width=0.4\textwidth]{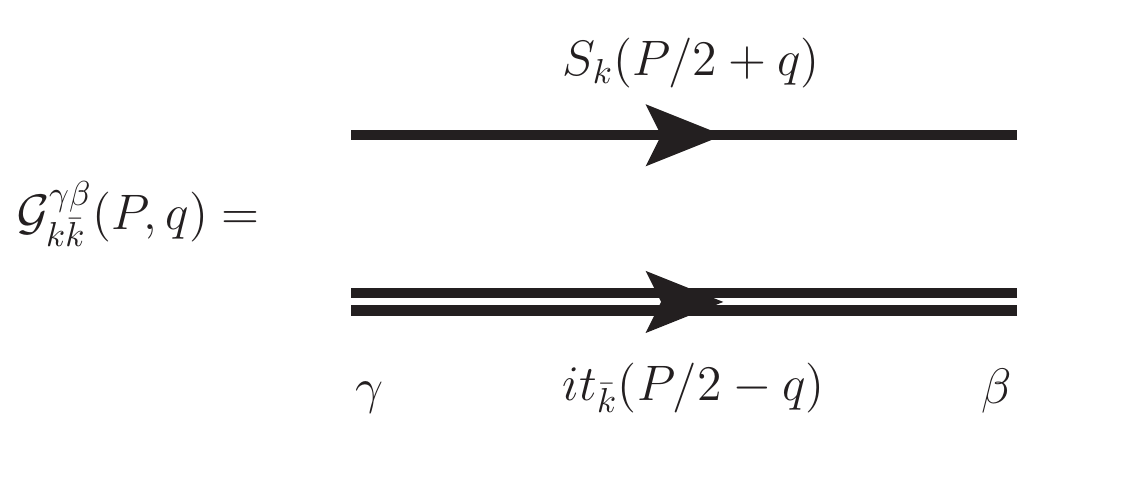}
  \end{center} 
  \caption{\label{fig:fadeev} Left panel: Effective coupling in the Fadeev equation which contains a quark exchange. Right panel: Two-particle (quark+diquark) propagator in the Fadeev kernel.}
\end{figure*}

The kernel reads~\cite{Buck:1992wz} 
\begin{equation}
\label{eq:kernel}
  L^{\bar{j}\bar{k} ,\alpha \beta}_{j k} (P^2,q,k) =  
  \ {\cal G}_{k \bar{k}}^{\gamma \beta} (P^2,q)  Z_{jk}^{\bar{k} \bar{j},\alpha \gamma} (q,k) \ ,  
\end{equation}
with a first term which accounts for the free quark and diquark propagators (see right panel of Fig.~\ref{fig:fadeev})
\begin{equation}
  {\cal G}_{k \bar{k}}^{\gamma \beta} (P^2,q) = S_k (P/2+q) \ it^{\gamma \beta}_{\bar{k}} (P/2-q) 
\end{equation}
and a second term
\begin{equation}
  Z_{jk}^{\bar{k} \bar{j},\alpha \gamma} (q,k)= \Omega_{jl}^{\bar{k},\gamma} \ S_l (-q-k) \ \Omega_{lk}^{\bar{j},\alpha} \ ,
\end{equation}
which represents an interaction with an exchanged quark (displayed in the left panel of Fig.~\ref{fig:fadeev}). 
We do not attempt here to justify the form of the Fadeev equation (\ref{eq:fadeev}) and we refer the reader to the original papers~\cite{Reinhardt:1992,Buck:1992wz} to learn the rigorous derivation and to learn more details.
 
\begin{figure*}[htp]
  \begin{center}
    \includegraphics[scale=0.5]{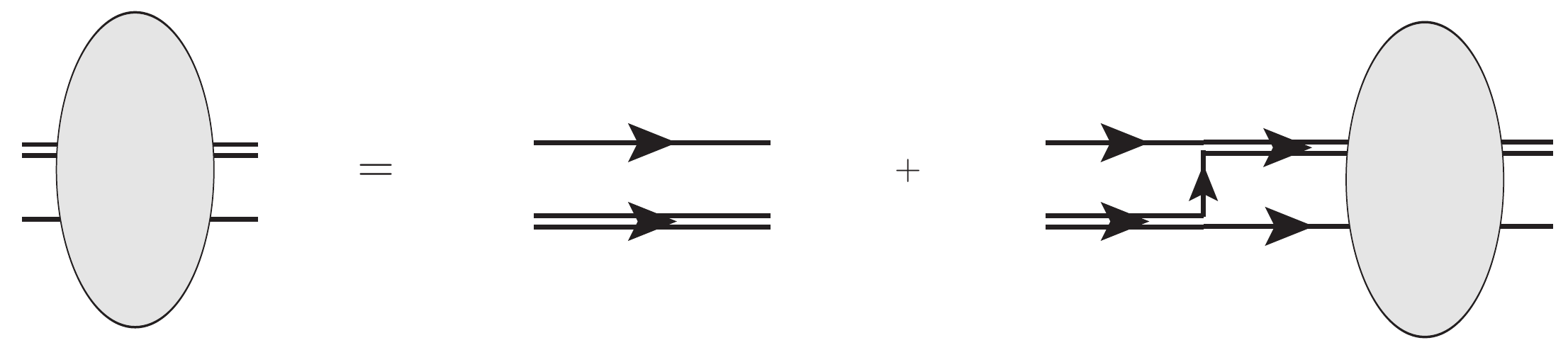}
  \end{center}
  \caption{\label{fig:motiv} Dyson equation for the baryon propagator ${\bf G}= {\cal G}  + {\cal G}  Z {\bf G}$.}
\end{figure*}

\begin{figure}[htp]
  \begin{center}
    \includegraphics[scale=0.42]{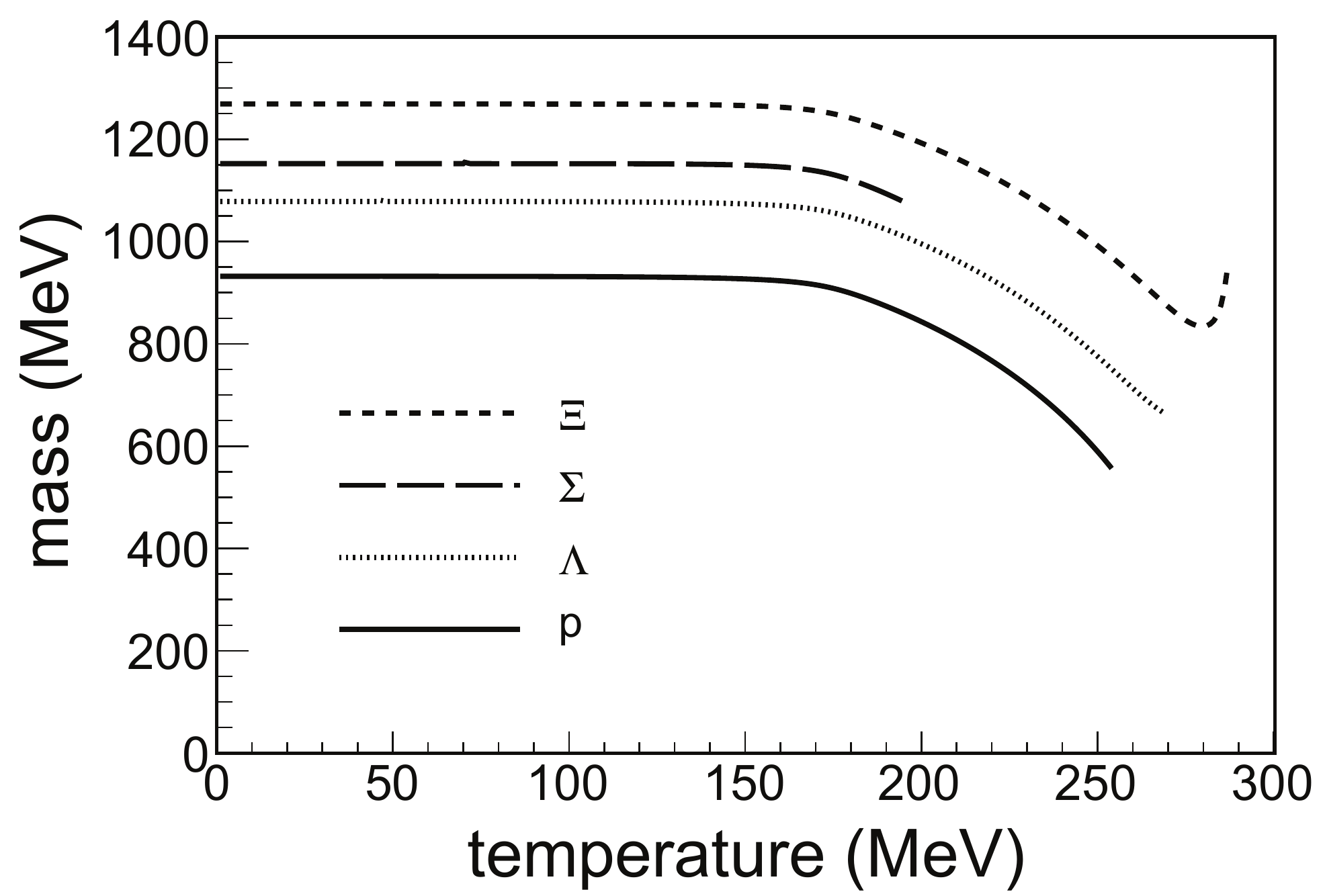}
    \includegraphics[scale=0.42]{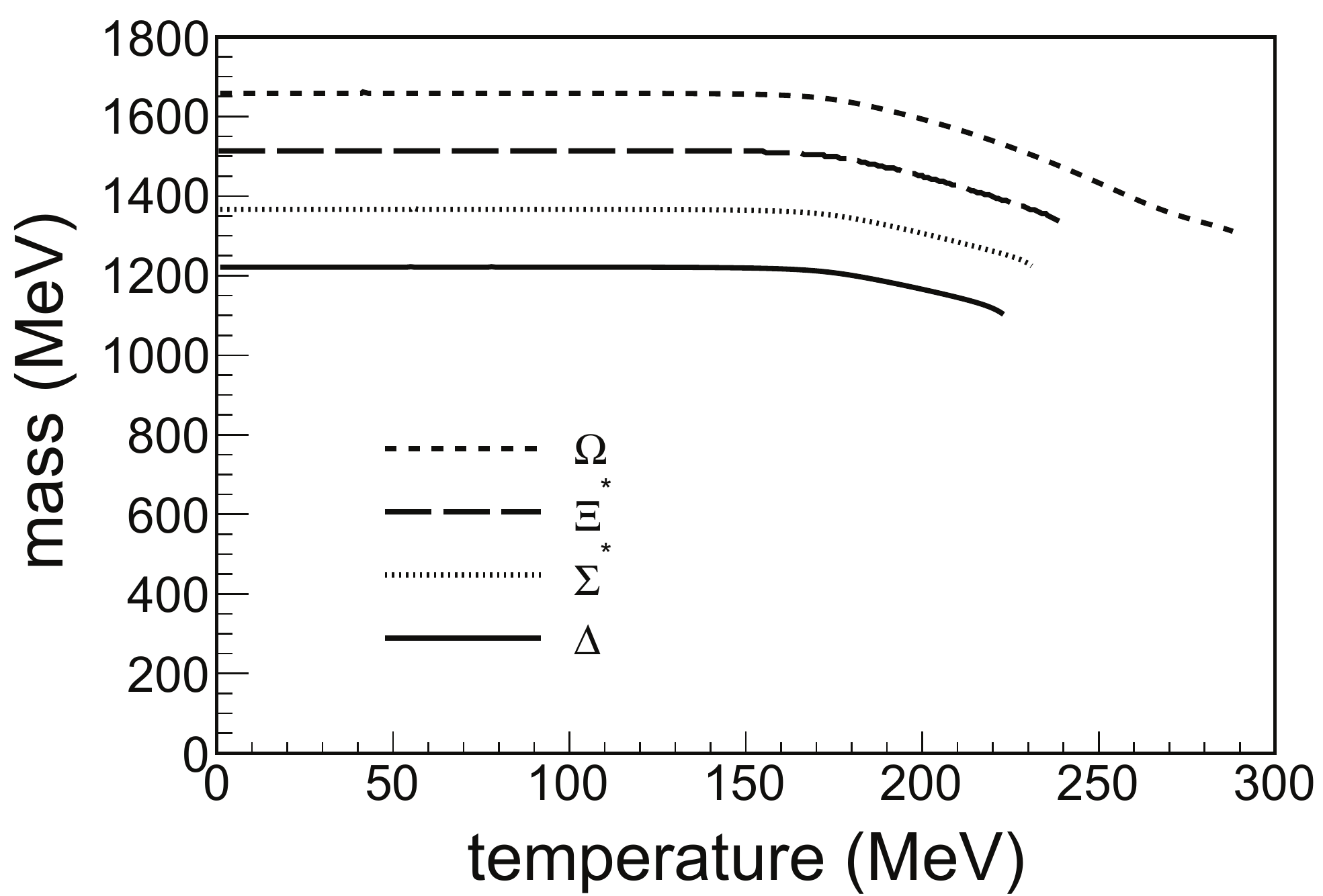}
  \end{center} 
  \caption{\label{fig:bar_pnjl} Baryon masses as a function of temperature for vanishing chemical potential in the PNJL model.}
\end{figure}

Nevertheless, we can provide a simple motivation for Eq.~(\ref{eq:fadeev}): If we denote by ${\bf G} (P^2)$ the full baryon propagator, one can form a Dyson equation by taking ${\cal G}$ as the leading order approximation (free propagation), and then considering $Z$ as the elementary interaction (see Fig.~\ref{fig:motiv}).
The Dyson equation can be symbolically written as
\begin{equation}
  {\bf G}= {\cal G}  + {\cal G}  Z {\bf G} \ ,
\end{equation}
whose solution reads
\begin{equation}
  {\bf G} = \frac{ {\cal G}}{1- {\cal G} Z} \ .
\end{equation}
The baryon masses are now extracted as the poles of the baryon propagator, so one needs to solve ${\bf G}^{-1} X (P^2=M^2_B)=0$, where $X$ is the baryon wave function. Explicitly,
\begin{equation}
  (1 - {\cal G} Z) X (P^2=M^2_B)= 0 \ ,
\end{equation}
at $P^2=M_B^2$, which is a simplified version of the more complete Eq.~(\ref{eq:fadeev}). Our results for the baryon masses at finite temperature  for vanishing chemical potential in both, the octet and decuplet, representations are shown in the PNJL model in Fig.~\ref{fig:bar_pnjl}.
We observe first of all the the ground state masses are quite well reproduced in the PNJL approach. Deviations from the experimental values are smaller than 5\%. This is astonishing in view of the fact that besides the parameters fixed already by the meson sector only one new parameter enters, the diquark coupling constant. We observe as well that with increasing temperature the baryon mass decreases and become finally unstable at the temperature where the lines end.  

\section{Relativistic Transport Theory}

We have seen that at high temperatures quarks and antiquarks are the essential degrees of freedom whereas at low energies the formation of mesons are thermodynamically preferred because the quark condensate becomes large at low temperature and hence the quarks become heavy. Mesons, on the contrary, become unstable at high temperature and develop a width due to their decay into quarks and antiquarks. The question we address in this section is how a finite size system develops, when we have brought it to a high temperature where the system is characterized by the quark degrees of freedom. In order to answer this question we need a transport theory. We could start out with a hydrodynamical approach but since we are interested not in the energy and particle density but in the processes by which quarks and antiquark pairs form mesons during the expansion and how the observables fluctuate in the vicinity of the phase transition, we search for a transport theory in which the time evolution of the quarks is addressed i.e. a molecular dynamics approach. The smallness of the quark mass at high temperature requires that this approach is covariant. 

\subsection{Relativistic molecular dynamics}

The starting point for the nonrelativistic equations of motion is the time evolution equation for a function $A$ defined on the $6N+1$-dimensional phase space:
\begin{equation}
  \frac{d A}{d t} = \frac{\partial A}{\partial t} + \{ A,\mathcal{H} \}  ,
\label{phsfct}
\end{equation} 
where  $\mathcal{H} ({\bf q},{\bf p})$ is the Hamiltonian of the system and $\{ A,B \}$ is the Poisson bracket of $A$ and $B$, defined as
 \begin{equation}
 \{ A,B \} = \sum_k^N   \frac{\partial A}{\partial {\bf q}_k}
                        \frac{\partial B}{\partial {\bf p}_k}
		              - \frac{\partial A}{\partial {\bf p}_k}
		                \frac{\partial B}{\partial {\bf q}_k} \ . 
\end{equation}
If we replace $A$ by either $\bf q$ or $ \bf p$ (which do not depend explicitly on time) we recover the Hamilton-Jacobi equations
\begin{equation}
  \frac{d {\bf q}}{d t} = \{ {\bf q} , \mathcal{H} \}
  = \frac{\partial \mathcal{H}}{\partial {\bf p} } \ ,
  \quad
  \frac{d {\bf p} }{d t} = \{ {\bf p} , \mathcal{H} \}
  =-\frac{\partial \mathcal{H}}{\partial {\bf q} } \ .
  \label{hameq}
\end{equation}
which give, for a  initial condition (${\bf q}_0,{\bf p}_0 $),  the desired trajectory of the particle in the phase space. For the later discussion it is important to note that Eqs.~(\ref{hameq}) are the differential equations for the trajectory in the $6N+1$ dimensional phase space on which the energy $\mathcal{H}$ is conserved.

Equation~(\ref{hameq}) cannot directly be made covariant containing the fourth component of the coordinate and of the energy-momentum 4-vector. Therefore, for the relativistic equations of motions we have to go back to the Poisson equations for 4-vectors
\begin{equation}
  \{ A,B \} = \sum_{k=1}^N
  \frac{\partial A}{\partial q_k^{\mu}}
  \frac{\partial B}{\partial p_{k \mu}} -
  \frac{\partial A}{\partial p_k^{\mu}}
  \frac{\partial B}{\partial q_{k \mu}} \ ,
\end{equation}
because in a dynamical system $q^\mu$ and $p^\mu$ depend on the time evolution parameter $\tau$, these quantities have to be taken at equal time $\tau$.

If the system is composed of several particles we find for the generators for the translation group
\begin{equation}
  P^\mu = \sum_k^N p_k^\mu \ ,
  \label{bigP}
\end{equation}
and for the Lorentz group [$SL (n=2,\mathbb{C}) \rightarrow \textrm{dim} = 2 (n^2 - 1) = 6$]
\begin{equation}
  M^{\mu\nu} = \sum_{k=1}^N q_k^\mu p_k^\nu - q_k^\nu p_k^\mu \ .
\end{equation}
These 10 generators define the Poincar\'e algebra:
\begin{equation}
  [P_\mu      , P_\nu         ] =   0, \quad ; \quad
  [M_{\mu \nu}, P_\rho        ] = g_{\mu \rho} P_\nu  - g_{\nu\rho} P_\mu ,
\end{equation}
\begin{equation}
  [  M_{\mu\nu} , M_{\rho\sigma} ]  =   g_{\mu\rho} M_{\nu\sigma} - g_{\mu\sigma} M_{\nu\rho}
  - g_{\nu\rho} M_{\mu\sigma} + g_{\nu\sigma} M_{\mu\rho}.
\label{poinal}
\end{equation}
The generator of a Poincar\'e transformation is given by
\begin{equation} 
  G = \frac{1}{2} \omega^{\mu\nu} M_{\mu\nu} - a^\mu P_\mu.
  \label{gmn}
\end{equation}
The space-time coordinates of the same event in two inertial frames ${\cal O}$ and ${\cal O}'$ are related by
\begin{equation}
  {q'}^\mu = q^\mu + \{q,G\}
           = q^\mu + \omega^\mu_\nu q^\nu + a^\mu
           =        \Lambda^\mu_\nu q^\nu + a^\mu.
  \label{poibra}
\end{equation}
World lines of particles are given by (${\bf q}_i(\tau),{\bf p}_i (\tau)$) and therefore physical trajectories (position and momentum of the particles as functions of the time $\tau$) have $6N+1$ dimensions. Thus we need \emph{constraints} to reduce the number of degrees of freedom in the relativistic phase space from $8N$ to $6N+1$. These constraints $K$ have to be Poincar\'e-invariant quantities.
\begin{equation} 
  \{K, M_{\mu\nu}\} = 0 , \quad \{K,P_\mu\}=0.
  \label{const4}
\end{equation}
In this lecture we demonstrate the idea and the consequence of this reduction for the case of one free particle. For the general case we have to refer to \cite{Marty:2012vs}. The trajectory in phase space on which a constraint $K$ is satisfied, is given by the solution of
\begin{eqnarray}
    \frac{d q^\mu(\tau)}{d \tau} &=& \lambda \{ q^\mu(\tau),K \}, \nonumber \\
    \frac{d p^\mu(\tau)}{d \tau} &=& \lambda \{ p^\mu(\tau),K \},
  \label{ham00}
\end{eqnarray}
with the initial condition $q(0)=q_0$ and $p(0)=p_0$. $\lambda$ is a Lagrange multiplier. In order to associate to each value of $\tau$ {\it one} point in phase space $( q(\tau), p(\tau))$ or, in other words, in order to create a world line a second constraint, $\chi(q^\mu,p^\mu,\tau)= 0$, has to be employed to fix $\lambda$. It relates the time  $q^0$ of the particle with a Lorentz-invariant system time $\tau$. The subspace we are interested in is determined by conserved $\chi$ and $K$ constraints. This is expressed by
\begin{equation} 
  \frac{d \chi}{d \tau} = \frac{\partial \chi}{\partial \tau}
  + \lambda \{ \chi(\tau),K \}=0 \to
 \lambda = - \frac{\partial \chi}{\partial \tau} \{\chi,K \}^{-1}.
\label{vcon}
\end{equation} 
To obtain the desired time evolution equations we replace $\chi$ in Eq.~(\ref{vcon}) by $q^\mu$ and $p^\mu$. To discuss the trajectory it is better to take a concrete example. Let be $K$ the mass shell constraint $K = p^\mu p_\mu - m^2 = 0$. The time constraint $\chi$  is much less self-evident. We investigate here the equations of motion for two choices of $\chi$: $\chi = q^0 - \tau = 0$ or $\chi = x_\mu p^\mu - m \tau = 0$. In the first case we obtain
\begin{equation}
  \frac{d q^\mu}{d \tau} = \lambda \{ q^\mu,K \} = \frac{p^\mu}{p^0} \ ,
  \label{ham112}
\end{equation}
whereas in the second
\begin{equation}
  \frac{d q^\mu}{d \tau} = \lambda \{ q^\mu,K \} = \frac{p^\mu}{m} \ .
  \label{ham113}
\end{equation}
In both cases we have $d p^\mu / d \tau =0$. These examples show the essential property of relativistic kinematics: In order to determine a trajectory in the 6N+1 dimensional phase space we have to impose constraints but these constraints are not unique. Different choices of the constraints yield a different time evolution of the system. Thus the time evolution in the 6N+1 dimensional phase space is not uniquely defined and other criteria like cluster separability have to be addressed.

This method can be extended to a system of $N$ interacting particles~\cite{Marty:2012vs}. In the (P)NJL approach the interaction is expressed by the change of the mass of the quarks, Eq.~(\ref{mesonm}). Assuming that the system is locally not far from equilibrium we calculate the local temperature of the environment in which the quark is localized. Knowing the temperature we can determine the masses. After having determined the masses of all particles we use the energy constraints to calculate the trajectories.  For a given initial condition and using the temperature and density dependent masses of the (P)NJL approach we can then follow the trajectories of all particles using Poincar\'e invariant kinematics.

\subsection{Collisional interactions}

\begin{figure}
  \begin{center}
  \includegraphics[width=0.35\textwidth]{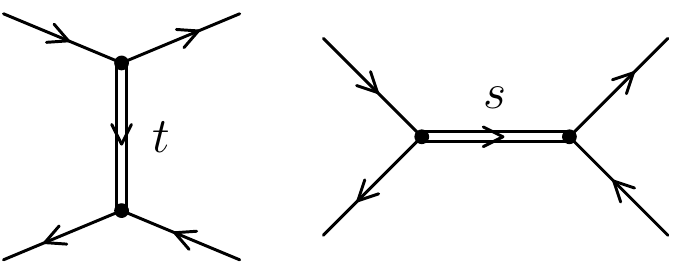}
  \includegraphics[width=0.35\textwidth]{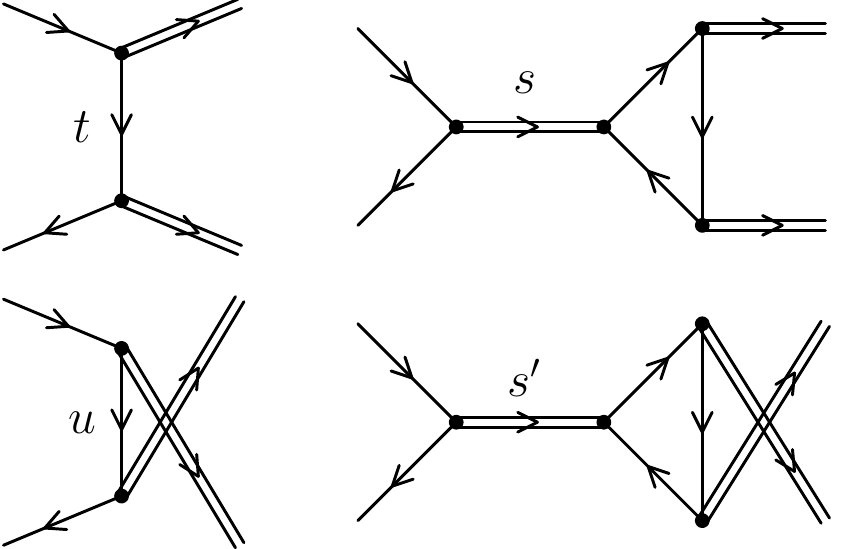}
  \label{feyndia}
  \end{center}
  \vskip -5mm
  \caption{Feynman diagrams for elastic $q \bar q$ scattering (left) and for hadronization (right).}
\end{figure}

\begin{figure}
  \begin{center}
    \includegraphics[width=0.35\textwidth]{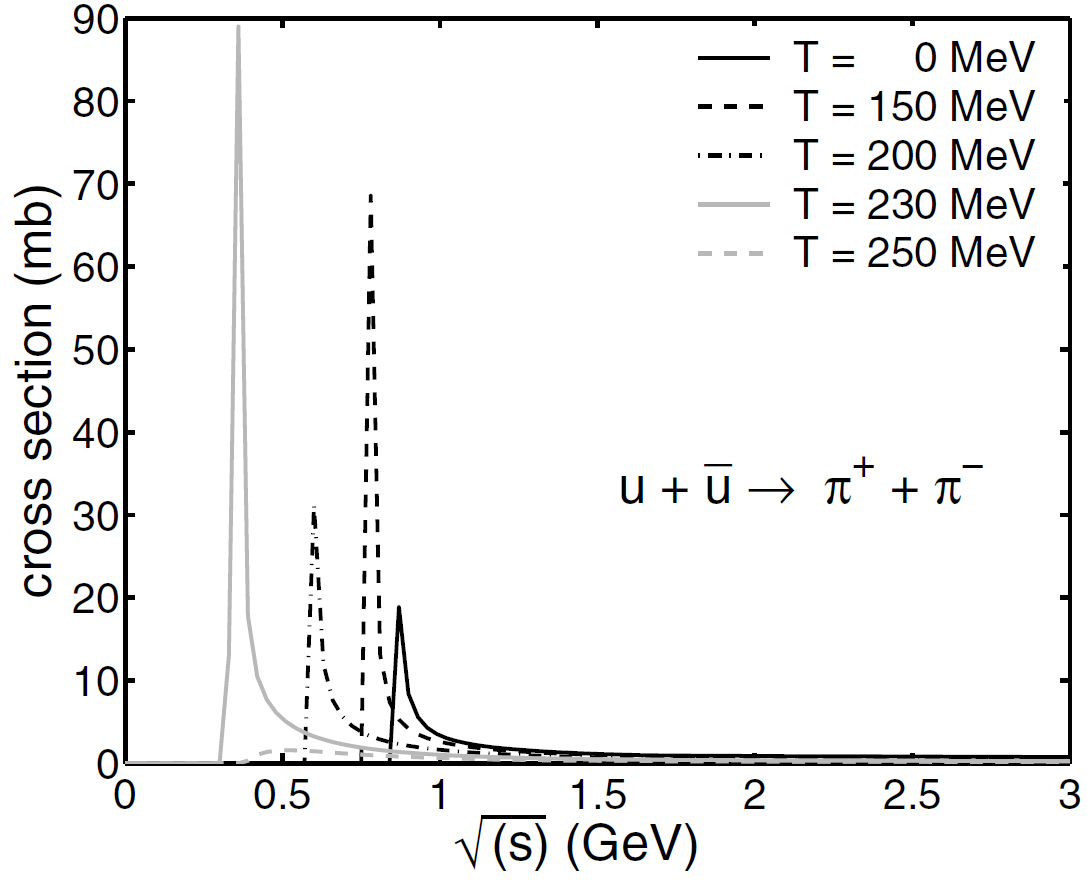}  
  \includegraphics[width=0.35\textwidth]{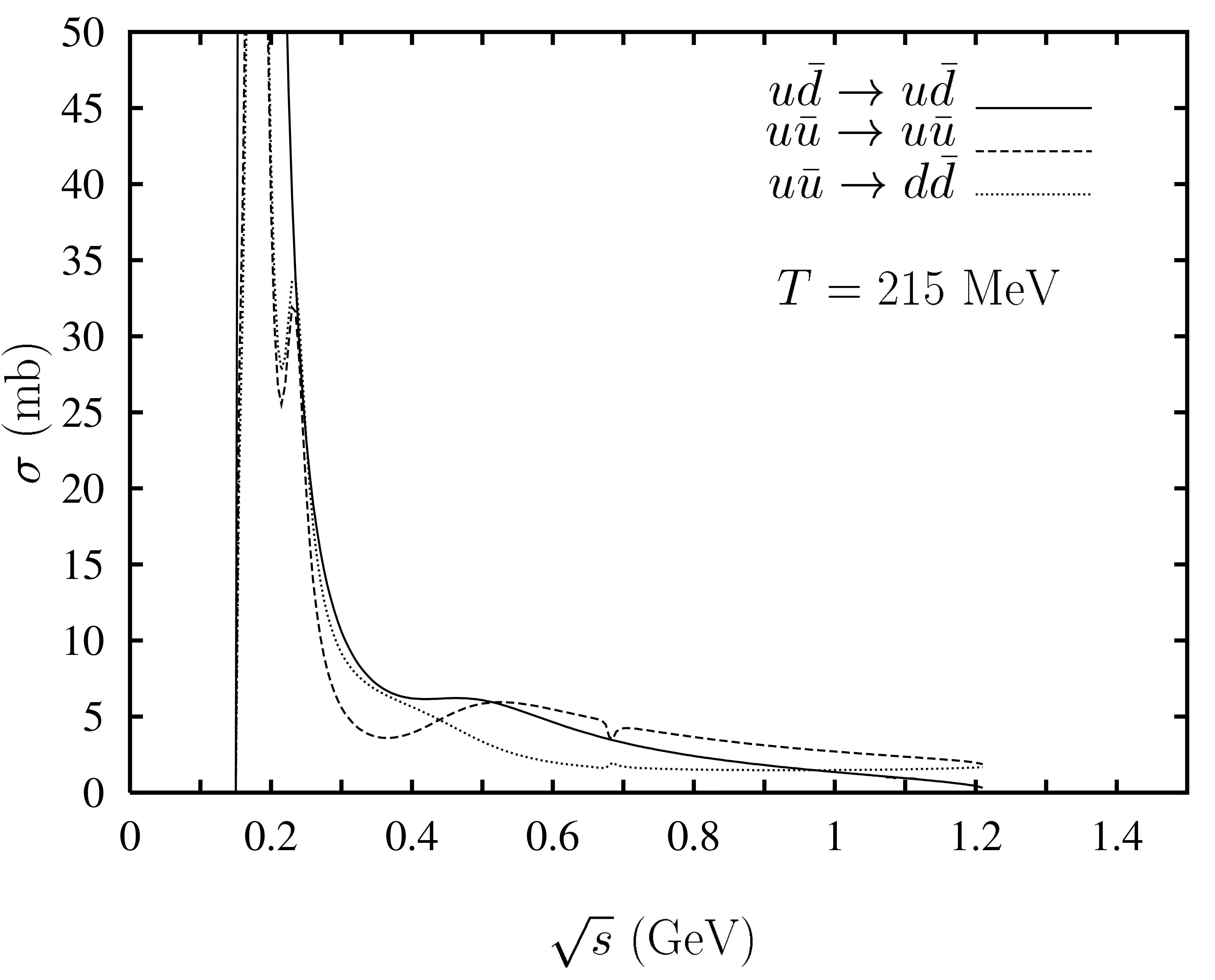}  
  \end{center}
  \vskip -5mm
  \caption{The cross section for the hadronization channel $u+\bar u \to \pi^+ + \pi^-$ (left) and the $q+\bar q \to q + \bar q$ cross section (right) as a function of the center of mass energy of the colliding quarks .\label{crosssec}}
\end{figure}

If created in heavy-ion collisions, the QGP will expand rapidly. Therefore, the cross sections between the constituents become dominant over the mean-field properties of the theory. In the NJL model these cross sections can be calculated via a $1 /N_c$  expansion of the Lagrangian \cite{Quack:1993ie}. All the details can be found in~\cite{Quack:1993ie,Rehberg:1995kh,Hufner:1994vd} and we refer to these references for details. There are two kinds of cross sections: cross sections with quarks or antiquarks  in the entrance and exit channels: $q(\bar q)+q(\bar q)\to q(\bar q)+q(\bar q)$ and hadronization cross sections $ q+ \bar q \to M+M$ where $M$ represents a meson. The Feynman diagrams for $q+\bar q \to q+\bar q$ collisions are given in the left side of Fig.~\ref{feyndia}, whereas in the right side of this figure we display the Feynman diagrams for hadronization cross sections. $s$ and $s'$ denote two $s$-channel diagrams with different intermediate mesons. The calculation of these diagrams requires the knowledge of the quark and meson masses and the vertex of the $q-\bar{q}-M$ interaction which we have already obtained when calculating the meson masses. Hence no new parameters are needed for calculating the cross sections.
The (P)NJL cross sections share a common property which is essential if we want to calculate with the help of the  potential and of the cross sections the time evolution of a system of quarks and anti-quarks. These cross sections have the typical size of perturbative-QCD cross sections,  a couple of milibarns, if we are away from the Mott temperature. Close to the Mott temperature, due to a resonant $s$-channel, the cross sections become huge and arrive at values of more than 100 mb. This is shown in Fig.~\ref{crosssec}. For an expanding plasma this means that early during the expansion --when the density is high-- a lot of collisions take place, but during the further expansion the rate decreases due to a decreasing density. If the system arrives at the Mott temperature the rate increases suddenly and elastic cross sections thermalize the system locally whereas the still larger hadronization cross sections convert the quarks into mesons. Finally, at temperatures below the Mott temperature we find a gas of mesons. The large hadronization cross sections provide an effective confinement in a theory which does not contain confinement.   

\section{Expansion of a quark-antiquark plasma and comparison with other theories and data}

\begin{figure}
  \begin{center}
  \includegraphics[width=0.35\textwidth]{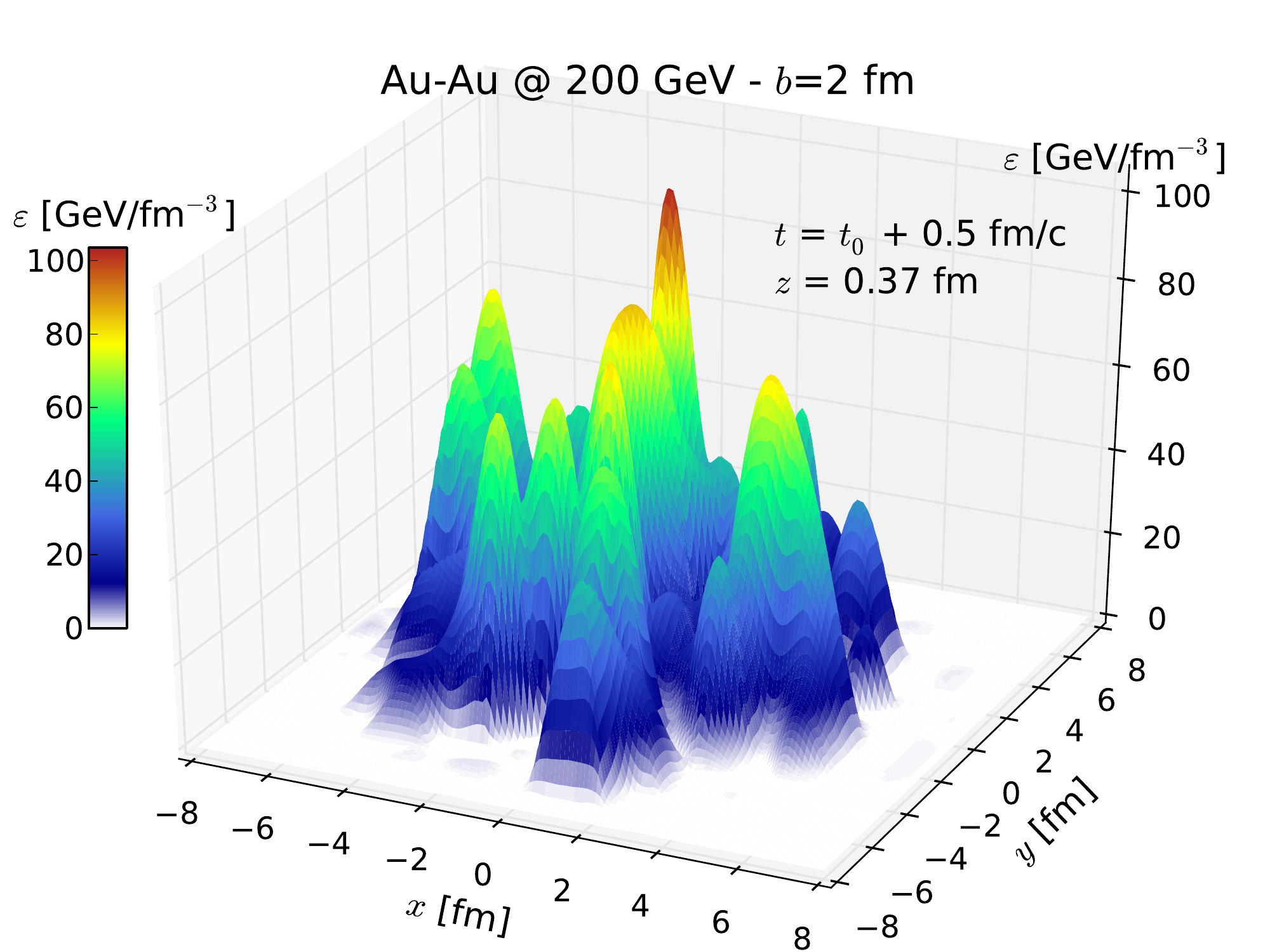}
  \includegraphics[width=0.35\textwidth]{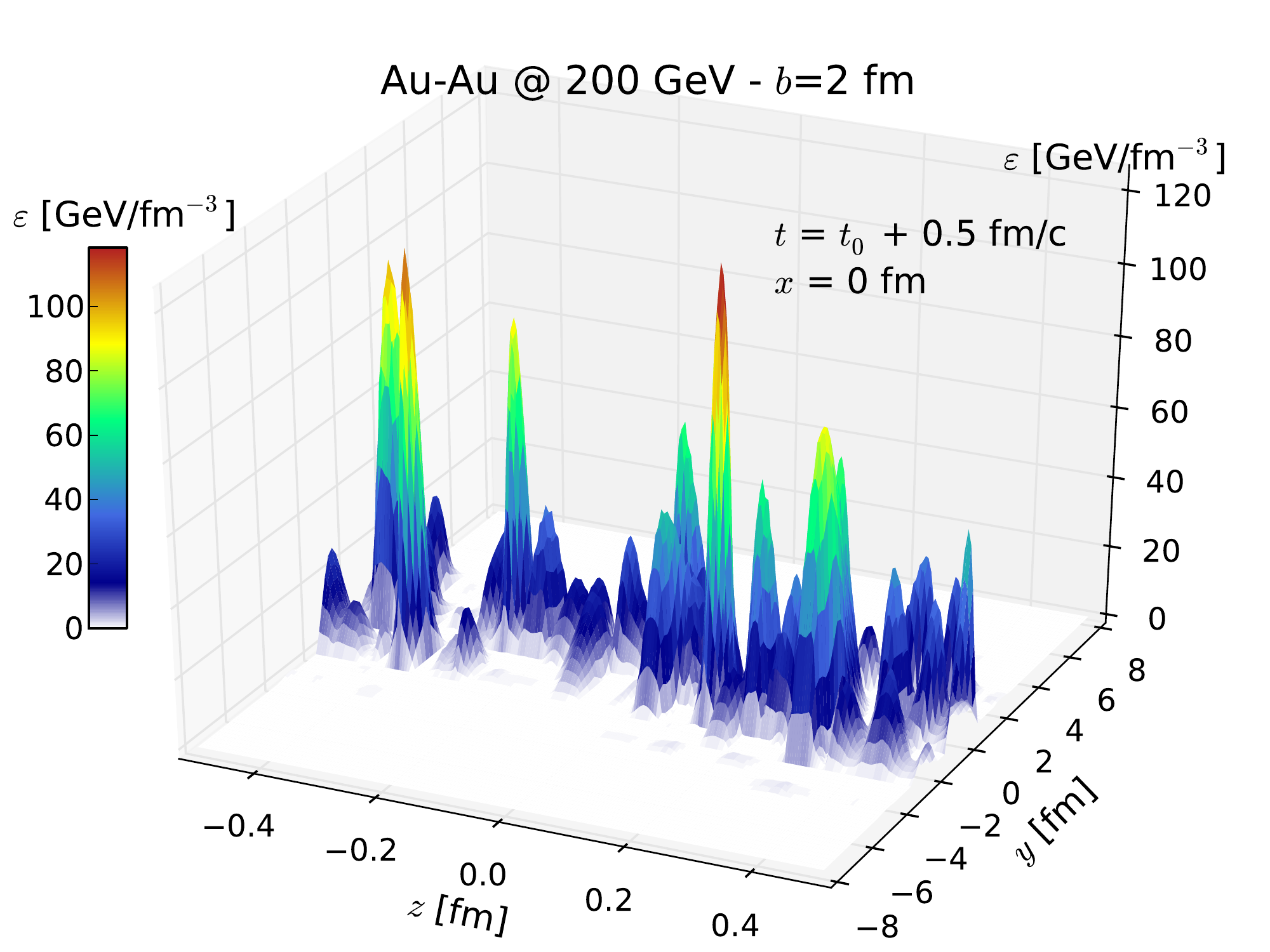}
  \end{center}
  \vskip -5mm
  \caption{Initial energy density $\epsilon$ of the expanding plasma as given by the PHSD initial condition. On the left panel the energy density in the $xy$ plane, on the right panel in the $zy$ plane. $z$ is the direction of the beam and $x$ the direction of the impact parameter.~\label{initen}}
\end{figure}

\begin{figure}
  \begin{center}
  \includegraphics[width=0.35\textwidth]{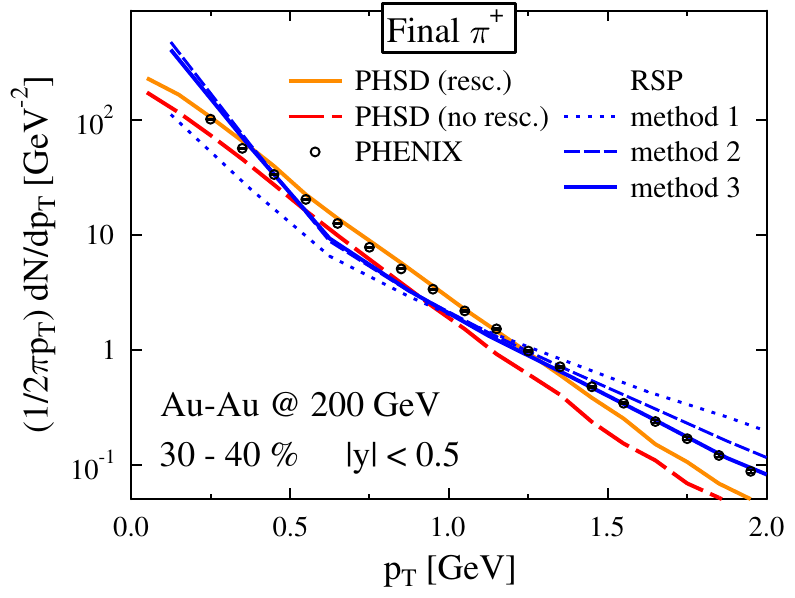}
  \includegraphics[width=0.35\textwidth]{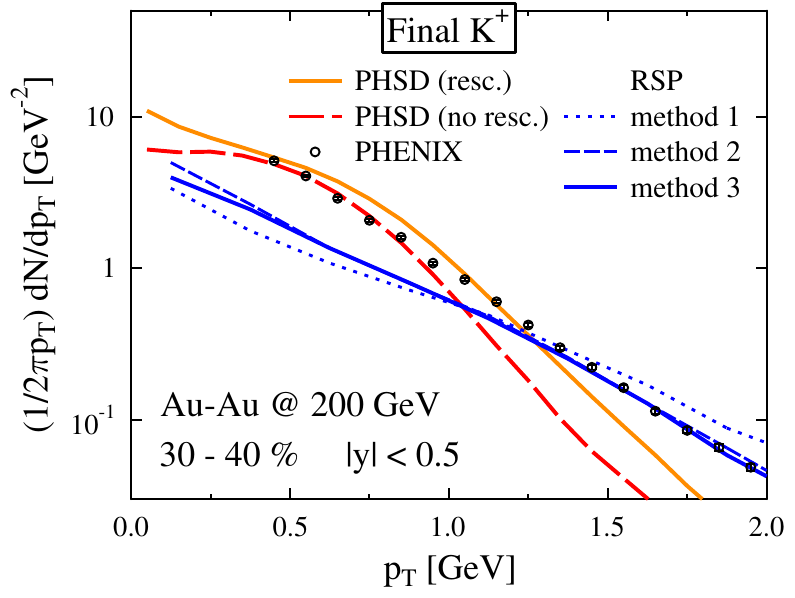}
  \end{center}
  \vskip -5mm
  \caption{The final $\pi^+$ (left) and $K^+$ (right) transverse momentum distribution of the NJL transport theory for AuAu with $\sqrt{s}$= 200 AGeV and a centrality of 30-40\% using the PHSD initial energy density distribution. Method 1-3 describe 3 different methods to convert this initial energy density distribution into the NJL quarks. Our results are compared with that of PHSD and with the experimental data.\label{resphsd}}
\end{figure}

In the preceding sections we have developed all the theoretical ingredients for a covariant transport theory based in the (P)NJL Lagrangian. In this section we present the results of the numerical realization of such a transport approach --dubbed RSP model (relativistic quantum molecular dynamics for strongly interacting matter with phase transition or cross-over)-- applied to the expansion of a quark+anti-quark plasma created in relativistic heavy-ion collisions~\cite{Marty:2012vs,Marty:2014zka}. The mechanism which brings the colliding nuclei in the time scale of 1 fm close to local equilibrium is not known yet and therefore we start our calculations at the time at which the highest density is reached (see Ref. \cite{Marty:2014zka} for details).  We use two initial conditions for the system at this time point: {\bf a)} either we assume a completely equilibrated system to perform model studies or {\bf b)} we use the initial condition of the PHSD model \cite{PHSD,Marty:2014zka} which has an anisotropic momentum distribution with a smaller component in transverse direction than in longitudinal direction and large energy density fluctuations. This initial energy density distribution is shown in Fig.~\ref{initen}. On the left panel we display the energy density distribution in the plane perpendicular to the beam direction, on the right panel in the $yz$ plane, where $z$ is the beam direction and $x$ the direction of the impact parameter.
In Fig. \ref{resphsd} we display the final distribution of $\pi^+$ and $K^+$ at the end of the expansion in a simulation of Au+Au reactions at $\sqrt{s} = 200 A$GeV and a centrality of 30-40\% with the PHSD initial condition. These results are compared with the results of PHSD calculations and the experimental data. The PHSD approach has been proven to describe a multitude of experimental data. Methods 1-3 characterize three methods to translate the PHSD initial energy density distribution into the quark degrees of freedom of the NJL approach. For details we refer to  \cite{Marty:2014zka}. We see that the  $\pi^+$ distribution is quite reasonable described, whereas for the $K^+$ we reproduce the high transverse momentum part but miss the yield at low momentum.  The origin is that in our approach the $ s+\bar s \to u(d) + \bar u(\bar d)$ cross section is higher than in PHSD, so for the same initial condition we form less $K^+$. 

\begin{figure}
  \begin{center}
   \includegraphics[width=0.35\textwidth]{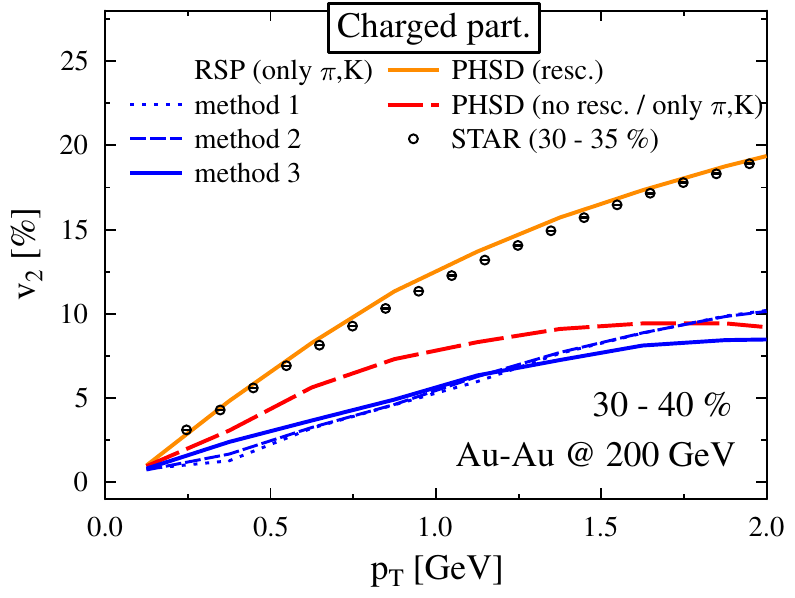}
  \end{center}
  \vskip -5mm
  \caption{Elliptic flow, $v_2$  produced in the NJL based transport approach (RSP) and in PHSD in comparison with the experimental data for AuAu, $\sqrt{s}= 200$ AGeV for a centrality of 30-40\%.\label{v2rhic}}
\end{figure}

\begin{figure}
  \begin{center}
  \includegraphics[width=0.35\textwidth]{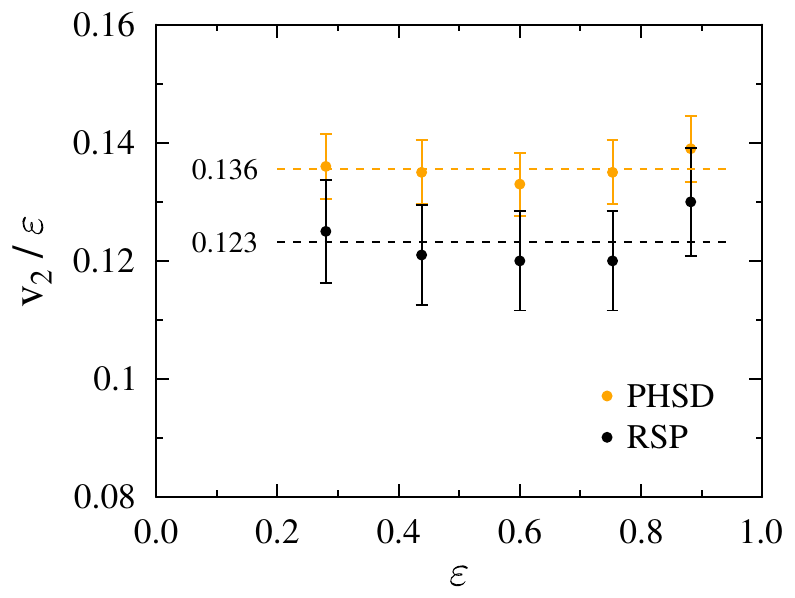}
  \includegraphics[width=0.35\textwidth]{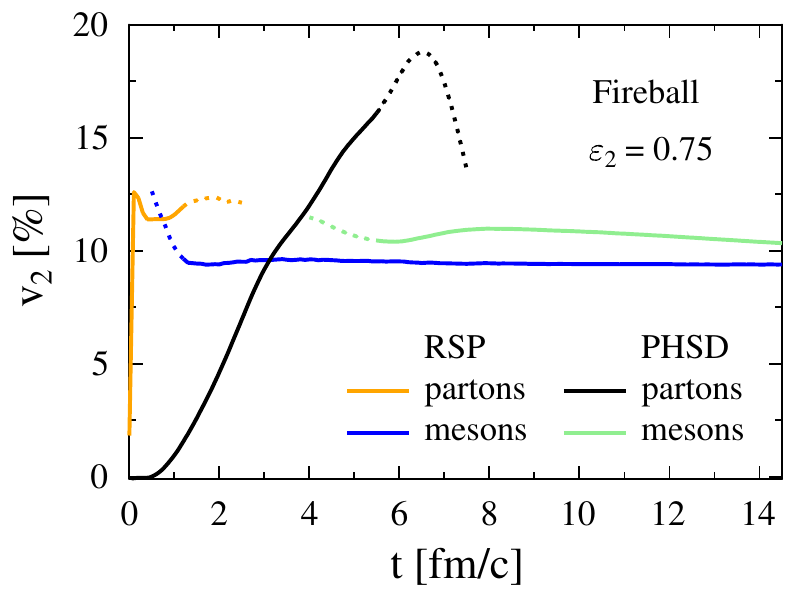}
  \end{center}
  \vskip -5mm
  \caption{Elliptic flow $v_2$ for a fireball initial condition (see text). On the right panel we display the centrality dependence of $v_2/ \epsilon$, for PHSD and the NJL based transport approach (RSP), on the right panel the time evolution of $v_2$ for the expanding fireball with an eccentricity of $\epsilon$ = 0.75.\label{v2fire}}
\end{figure}

Another observable of interest is the elliptic flow. In a hydrodynamical approach to describe the expanding plasma a elliptic flow is expected because the almond shape form of the overlap region between projectile and target is transformed into a larger pressure in the reaction plane that out of the reaction plane. In ideal hydrodynamics we expect that $v_2/ \epsilon$, where $\epsilon$ is the eccentricity of the overlap region and $v_2$ the second coefficient in the expansion $\frac{dN}{d\phi}=C(1+2v_1\cos{\phi}+2v_2\cos{2\phi}+..)$ of the azimuthal distribution of particles, is independent of the impact parameter. Since none of the two transport  approaches which we discuss here is of hydrodynamical nature the $v_2$ they produce cannot be deduced from general considerations. Fig. \ref{v2rhic} displays the results as compared to the experimental data. We see that PHSD agrees with the data. The NJL transport approach does not contain hadronic rescattering yet and therefore we have to compare its result with PHSD without rescattering, which is presented by the red line. Thus without rescattering we find good agreement between the NJL transport approach and PHSD.  
In order to study a bit more the expansion of the plasma we perform calculations in which we use a simplified initial condition, an expanding fireball whose geometry is given by the overlap of projectile and target. Fig. \ref{v2fire}, left,  shows that for such an initial condition the results for both transport approaches are compatible with hydrodynamical calculations, means $v_2/\epsilon$ is independent of the centrality of the interaction and that the $v_2$ value of both approaches differ by only 10\%. The dynamical creation of this $v_2$ value is, however, very different in the approaches. This is shown on the right hand side. The almost massless partons in the NJL approach produce the $v_2$ very fast and the hadronization lowers the $v_2$ value only little whereas in the PHSD approach where the partons are heavy the expansion as well as  the creation of $v_2$ is much slower. The hadronization lowers the $v_2$ by 30\%. So quite different scenarios for the expansion of the plasma give rather similar values for the final $v_2$. 

\begin{figure}
  \begin{center}
   \includegraphics[width=0.5\textwidth]{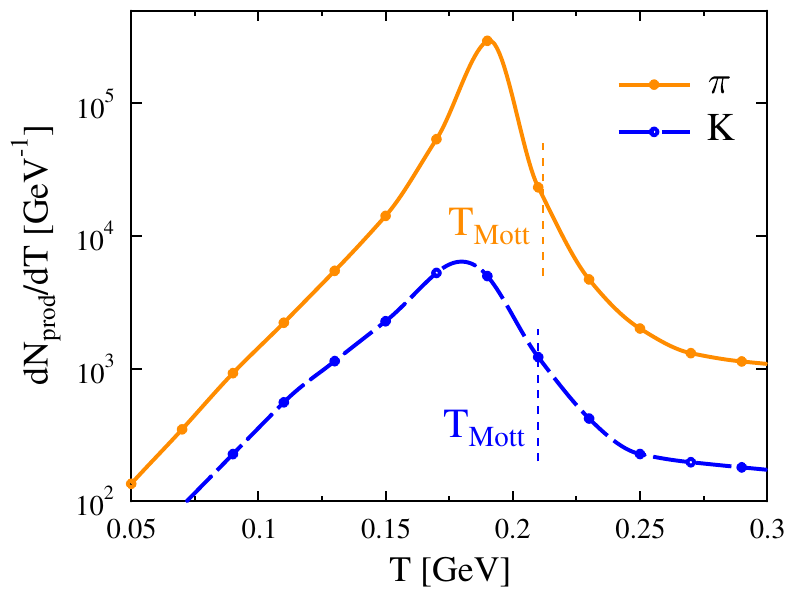}
  \end{center}
  \vskip -5mm
  \caption{Local temperature distribution around the  $q\bar q$ pair at the moment in which it creates pions which are finally observed.\label{pitemp}}
\end{figure}

It is interesting to study when the finally observed mesons are produced. As discussed the mesons get stable at temperatures below the Mott temperature. In an expanding plasma this transition from quarks to hadrons is not that sudden. We observe a broad distribution as seen in Fig.\ref{pitemp}. Mesons created above the Mott temperature can survive the time in which the local temperature passes the Mott temperature and mesons can be created well below the Mott temperature if only then the $q$ and $\bar q$ come sufficiently close to form mesons and the reaction becomes exothermic and hence favoured. 

\section{Summary}

In summary, we have presented in this lecture the whole way from the definition of the (P)NJL Lagrangian to a transport approach based on this Lagrangian which can be compared with experimental data and other theories. The transport approach does not need any additional input as compared to the (P)NJL Lagrangian whose parameters are fixed by vacuum values of meson masses and decay constants. For a given initial condition for the expansion of a quark-antiquark plasma, which we borrow from the PHSD approach because it can presently not be calculated, in this transport approach the predictions for the observables are close to the experimental results (and close to that obtained in the PHSD approach). Future steps will be to include baryons in the transport approach and to include the interaction between gluons, presented by the Polyakov loop and the quarks, which has recently been calculated in a Dyson-Schwinger approach. This interaction will bring the equation of state of the P(NJL) approach close to that of recent lattice gauge calculations.  

\vspace*{0.3cm}
This work was supported by the LOEWE center "HIC for FAIR". JMTR is supported by the Program TOGETHER from Region Pays de la Loire and the European I3-Hadron Physics program. The computational resources have been provided by the LOEWE-CSC.

\end{document}